\documentclass[aps,preprint]{revtex4}
\usepackage[dvips]{graphicx}
\usepackage{amssymb}
\usepackage{color}

\usepackage[normalem]{ulem}
\newcommand{\soutr}{\bgroup\markoverwith{\textcolor{red}{\rule[.5ex]{2pt}{1pt}}}\ULon}

\begin{document}
\draft

\title{Non-Weyl microwave graphs}

\author{Micha{\l} {\L}awniczak,$^{1}$ Ji\v{r}\'{\i} Lipovsk\'{y},$^{2}$ and Leszek Sirko$^{1}$}
\address{$^{1}$Institute of Physics, Polish Academy of Sciences, Aleja  Lotnik\'{o}w 32/46, 02-668 Warszawa, Poland\\
$^{2}$Department of Physics, Faculty of Science, University of Hradec Kr\'alov\'e, Rokitansk\'eho 62, 500 03 Hradec Kr\'alov\'e, Czechia\\
}
\date{\today}

\begin{abstract}
One of the most important characteristics of a quantum graph is the average density of resonances, $\rho = \frac{\mathcal{L}}{\pi}$, where  $\mathcal{L}$ denotes the length of the graph. This is a very robust measure. It does not depend on the number of vertices in a graph and holds also for most of the boundary conditions at the vertices. Graphs obeying this characteristic are called \emph{Weyl} graphs. Using microwave networks which simulate quantum graphs we show that there exist graphs which do not adhere to this characteristic. Such graphs will be called \emph{non-Weyl} graphs. For standard coupling conditions we demonstrate that the transition from a \emph{Weyl} graph to a \emph{non-Weyl} graph occurs if we introduce a balanced vertex. A vertex of a graph is called balanced if the numbers of infinite leads and internal edges meeting at a vertex are the same. Our experimental results confirm the theoretical predictions of [E.~B. Davies and A. Pushnitski, Analysis and PDE {\bf 4}, 729 (2011)] and are in excellent agreement with the numerical calculations yielding the resonances of the networks.
\end{abstract}


\bigskip
\maketitle

\section{Introduction}
The behavior of a quantum particle on a physical network can be described by a model of a quantum graph  \cite{Pauling36, Ruedenberg53, Exner88}, which has been extensively developed  in the last thirty years \cite{Kottos1997, Blumel2002, BK, Pluhar2014}. In this model, the quantum Hamiltonian acting as the negative second derivative with appropriate coupling conditions at the vertices is introduced. The most physical ones are the standard coupling conditions, also called the Neumann boundary conditions, which prescribe the continuity of the functional value and vanishing of the sum of outgoing derivatives at each vertex. For more details on quantum graphs, we can refer the reader to the book \cite{BK} and the references therein.
Quantum graphs were used  to simulate, e.g., mesoscopic quantum systems \cite{Kowal1990,Imry1996}, quantum wires \cite{Sanchez1988}, and optical waveguides \cite{Mittra1971}.

 Quantum graphs can be modeled by microwave networks since both systems are described by the same equation, namely, it was demonstrated that
the telegrapher's equation for  microwave networks is formally
equivalent to the one-dimensional Schr\"odinger equation describing quantum graphs \cite{Hul2004,Sirko2016}. Since the introduction of microwave networks they were successfully used, e.g., to extend a famous question asked by Mark Kac "Can one hear the shape of a drum", posed originally in the context of dissipationless isospectral systems, to scattering systems such as isoscattering ones \cite{Hul2012}, and to demonstrate the power of missing level statistics in variety of applications \cite{Bialous2016}.

There are two ways how to define resonances in quantum graphs. The resolvent resonances are poles of the resolvent continued to the second Riemann sheet. The scattering resonances are poles of the continuation of the determinant of the scattering matrix. It has been proven \cite{EL1,Li5} that the set of resolvent resonances is equal to the union of the set of scattering resonances and the set of eigenvalues with eigenfunctions supported on the compact part of the graph. Choosing the ratios of the lengths of the graph to be irrational, one can assure that there are no eigenvalues with compactly supported eigenfunctions and therefore the resolvent resonances can be ``seen'' via the singularities of the determinant of the scattering matrix.

 The fundamental question is what the number of resonances is. For a compact graph without infinite leads, according to the
definition of resolvent resonances (Definition 1.1 in Ref. \cite{DP}), the set of resonances would coincide
with the eigenvalues. In this case, the counting function of the number of eigenvalues with $k$ in the interval (0,R) ($k$ is the square root of energy) satisfies the Weyl's law \cite{DP,Weyl11}
\begin{equation}
  N(R) = \frac{\mathcal{L}}{\pi}R + \mathcal{O}(1)\,,\label{eq:asym}
\end{equation}
 where $\mathcal{L}$ is the sum of the lengths of the edges of the graph and $\mathcal{O}(1)$ is a function which in the limit $R\rightarrow +\infty $ is bounded by a constant of order of 1. In the case of a non-compact graph one would expect that the same asymptotics would hold also for the number of resolvent resonances
  in the complex $k$-plane that lie within a semicircle with $\mathrm{Re\,}k>0$ of radius $R$ centered at the origin,
 only with $\mathcal{L}$ denoting the sum of the lengths of the internal edges of the graph. As pointed out by Davies and Pushnitski \cite{DP}, this is not the case for all the graphs. There exist graphs for which the coefficient by the leading term of the asymptotics is smaller than expected. We will call these graphs \emph{non-Weyl}. The graphs satisfying the asymptotics (\ref{eq:asym}) are called \emph{Weyl} graphs.

In \cite{DP} a simple geometric condition for graphs with standard coupling is proved, which distinguishes Weyl and non-Weyl graphs. The graph is non-Weyl if and only if there exists a vertex for which the number of internal and external edges (infinite leads) is the same. In \cite{DEL} the condition distinguishing non-Weyl graphs was studied for general coupling conditions. The paper \cite{Li6} finds bounds on the coefficient by the leading term of the asymptotics for a non-Weyl graph. In the paper \cite{EL3} it was found that the presence of the magnetic field cannot change a non-Weyl graph into the Weyl one but it can change the coefficient by the leading term of a non-Weyl graph.

In the present paper, we experimentally verify the geometric condition introduced by Davies and Pushnitski by constructing two similar microwave networks, possessing the same lengths of the internal edges -- one Weyl and the other one non-Weyl -- and measuring their scattering matrices.
However, one should stress out that dynamics of a charged particle in the presence of the magnetic field will not be considered in our experimental and theoretical analyses.

\section{Simulations of quantum graphs using microwave networks}\label{sec2}

Quantum graphs are often considered as idealizations of physical networks in the
limit where the lengths of the wires are much bigger than their widths. They
were successfully applied to model a broad range of physical problems, see, e.g.,
\cite{Gnutzmann2006}. Quantum graphs can also be realized
experimentally. Using contemporary epitaxial techniques it is possible to
design and fabricate quantum nanowire networks \cite{Samuelson2004,Heo2008}.

In a seminal paper by Hul et al.\ \cite{Hul2004} it was shown that quantum graphs
can be successfully simulated by microwave networks comprising microwave junctions and coaxial cables.  In the present investigations the  SMA-RG402 coaxial cables were used. The SMA-RG402 cable consists of an inner conductor of radius $r_1=0.05$~cm, which is surrounded by a concentric conductor of inner radius $r_2=0.15$~cm. The space between them is filled with Teflon with a dielectric constant $\varepsilon\simeq 2.06$. Below the cut-off frequency of the TE$_{11}$ mode $\nu_{c}\simeq\frac{c}{\pi (r_1+r_2)\sqrt{\varepsilon}} \simeq 33$~GHz~\cite{Savytskyy2001}, where $c$ is the speed of light in vacuum, only the fundamental TEM mode can propagate inside a coaxial cable. It is important to point out, that not the geometric lengths $\ell^g_i$ of the coaxial cables, but the optical lengths $\ell_i=\ell^g_i\sqrt{\varepsilon}$ yield the lengths of the edges in the corresponding quantum graph.

Therefore,
 properties of quantum graphs can be studied experimentally using
microwave networks with the same topology and boundary conditions at the
vertices. They provide an extremely rich system for the experimental and the theoretical study of quantum systems, that exhibit a chaotic dynamics in the classical limit. A broad range of spectral and scattering properties of microwave networks have
been studied in Refs. \cite{Hul2004,Hul2005,Lawniczak2008,Lawniczak2010,Sirko2016,Bialous2016,Stockmann2016,Dietz2017}. It is important to point out that only microwave networks allow for simulations of variety of quantum chaotic systems whose spectral properties can be described by the three main symmetry classes:  Gaussian orthogonal ensemble (GOE) \cite{Hul2004,Sirko2016}, Gaussian unitary ensemble (GUE) \cite{Hul2004,Lawniczak2010,Bialous2016} and Gaussian symplectic ensemble (GSE) \cite{Stockmann2016} in the Random Matrix Theory.

A finite compact quantum graph consists of $V$ vertices connected by $B$ edges. Each vertex
$i$ of a graph is connected to the other vertices by $b_{i}$ edges, $b_{i}$ is
called the valency of the vertex $i$. A wave function propagates on each edge
of a graph according to the one-dimensional Schr\"odinger equation. The
properties of a graph are determined by the lengths of edges connecting
vertices and vertex boundary conditions relating amplitudes of the waves
meeting at each vertex. Here we consider graphs with the standard boundary conditions which
impose the continuity and vanishing of the sum of the derivatives calculated at
a vertex $i$ of waves propagating in edges meeting at $i$. Such graphs are standard examples of Weyl graphs.

In order to study the transition from Weyl graphs to non-Weyl graphs we have to introduce additional elements of a graph - infinite leads which can be connected to some subset of vertices of a compact graph. The transition from a Weyl graph to a non-Weyl graph occurs if we introduce a balanced vertex. A vertex $j$ of a graph is called balanced if the number of leads attached to $j$ equals the number of internal edges attached to $j$. If the number of leads attached to a vertex is smaller or greater than the number of attached internal edges, the vertex will be called unbalanced.

In the case of a non-Weyl graph possessing a single balanced vertex the number of resonances is given by the formula \cite{DP}

\begin{equation}
  N_{nW}(R) = \frac{\mathcal{L'}}{\pi}R + \mathcal{O}(1)\,,\label{eq:non_Weyl_asym}
\end{equation}
where $\mathcal{L'}=\mathcal{L}- \ell_s$ is the effective size of a non-Weyl graph and $\ell_s$ is the length of the shortest edge emanating from the balanced vertex.

In order to test experimentally properties of the Weyl and non-Weyl graphs we consider two graphs shown in Fig.~1(a) and Fig.~1(b). Both graphs are characterized by the same length $\mathcal{L}=\sum_{i=1}^7 \ell_i$.
 The Weyl graph in Fig.~1(a) contains two unbalanced vertices containing infinite leads $L^{\infty}_1$ and $L^{\infty}_2$. The non-Weyl graph in Fig.~1(b) contains one balanced vertex with two attached leads $L^{\infty}_1$ and $L^{\infty}_2$. The leads are marked with red broken lines. As discussed above we expect that the non-Weyl graph in Fig.~1(b) has smaller number of resonances.

 Two corresponding microwave networks constructed from microwave
coaxial cables are shown in Fig.~1(c) and Fig.~1(d).

\begin{figure}[tb]
\includegraphics[width=0.8\linewidth]{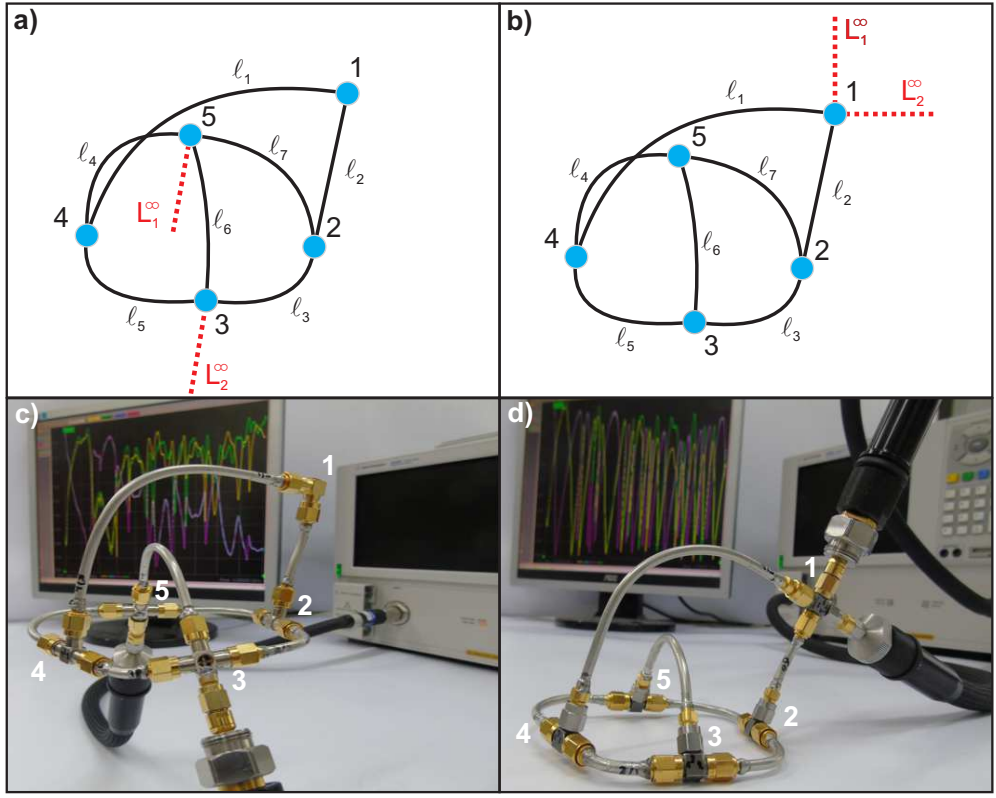}
\caption{
Panels (a) and (b) show the schemes of  Weyl and non-Weyl graphs, respectively. Both graphs are characterized by the same length $\mathcal{L}=\sum_{i=1}^7 \ell_i$.  The Weyl graph contains two unbalanced vertices containing infinite leads $L^{\infty}_1$ and $L^{\infty}_2$. The non-Weyl graph contains one balanced vertex with two attached leads $L^{\infty}_1$ and $L^{\infty}_2$. The leads are marked with red broken lines.
Panels (c) and (d) show the corresponding  Weyl and non-Weyl microwave networks constructed from microwave
coaxial cables and joints. The optical lengths of the networks are the same and are equal to $\mathcal{L}$.  The microwave networks are connected to the VNA with the two elastic microwave cables which is equivalent to attaching of
two infinite leads $L^{\infty}_1$ and $L^{\infty}_2$ to quantum graphs.
}
\label{Fig1}
\end{figure}

We prove that the effective size of the non-Weyl graph is $\mathcal{L'} = \mathcal{L}-\ell_s$,  where $\ell_s=\ell_2$ is the length of the shortest edge emanating from the balanced vertex 1 in Figs.~1(b) and 1(d). Let us introduce a fictitious vertex of valency two with standard coupling at the edge (1,4) at the distance $\ell_2$ from the vertex 1 and denote it by 6. We denote the wavefunctions on the edges (1,6) and (1,2) by $u_1(x)$ and $u_2(x)$, respectively, with $x=0$ at the vertex~1. Similarly, we denote the wavefunctions on the leads $L^{\infty}_1$ and $L^{\infty}_2$ by $f_1(x)$ and $f_2(x)$, again with $x=0$ at the vertex~1. The coupling condition at the vertex~1 then yields
\begin{equation}
\label{eq:coupling_cond}
  u_1(0) = u_2(0) = f_1(0) = f_2(0)\,,\quad u_1'(0)+ u_2'(0)+ f_1'(0)+ f_2'(0) = 0\,.
\end{equation}
Now we introduce symmetrization and antisymmetrization of the previously defined components of wavefunctions
\begin{equation}
\label{eq:sym_desym}
v_+ = \frac{1}{\sqrt{2}}(u_1+u_2),\quad v_- = \frac{1}{\sqrt{2}}(u_1-u_2),\quad
g_+ = \frac{1}{\sqrt{2}}(f_1+f_2),\quad g_- = \frac{1}{\sqrt{2}}(f_1-f_2).
\end{equation}
From the coupling conditions at the vertex 1 it follows using $u_1(0) = u_2(0)$ and $f_1(0) = f_2(0)$ that
\begin{equation}
\label{eq:sym_desym_2}
\begin{array}{l}
  v_+(0) = \frac{1}{\sqrt{2}}(u_1(0)+u_2(0)) = \sqrt{2}\, u_1(0),\quad g_+(0) = \frac{1}{\sqrt{2}}(f_1(0)+f_2(0)) = \sqrt{2}\, f_1(0)\,,\\
  v_-(0) = \frac{1}{\sqrt{2}}(u_1(0)-u_2(0)) = \frac{1}{\sqrt{2}}(u_1(0)-u_1(0)) = 0\,,\\
  g_-(0) = \frac{1}{\sqrt{2}}(f_1(0)-f_2(0)) = \frac{1}{\sqrt{2}}(f_1(0)-f_1(0)) = 0\,.
\end{array}
\end{equation}
The coupling condition can be in the new functions written (using $u_1(0) = f_1(0)$) as
\begin{equation}
\label{eq:sym_desym_3}
  v_+(0) = g_+(0)\,,\quad v_+'(0)+g_+'(0) = 0 \,, \quad v_-(0) = g_-(0) = 0\,.
\end{equation}

Notice that the condition between the symmetric functions $v_+$ and $g_+$ is the standard condition connecting an internal edge and a lead, while in the antisymmetric subspace we have Dirichlet condition. We denote by $h$ the wavefunction component on the rest of the graph (edges (2,3), (2,5), (3,4), (3,5), (4,5) and (4,6)). Then the map
\begin{equation}
\label{eq:map}
  U : (u_1,u_2,f_1,f_2, h)^\mathrm{T} \mapsto (v_+,v_-,g_+,g_-, h)^\mathrm{T}
\end{equation}
is unitary and transforms the ``old'' Hamiltonian $H$ for the graph in Fig.~\ref{Fig1}(b) to the ``new'' Hamiltonian $H_U = U H U^{-1}$. The graph for the Hamiltonian $H_U$ connects an internal edge
of length $\ell_2$ with an external lead by standard condition (continuity of the function and its derivative). Hence there is no interaction at this vertex (the vertex of valency two with the standard condition can be removed) and these two edges may be replaced by one external lead, thus reducing the effective size of the graph by $\ell_2$. There will arise a new, more complicated, coupling condition at the real vertex 2 and the fictitious vertex 6 which joins these two vertices and can be given by the Proposition~7.1 in Ref.\cite{DEL}. This condition assures that the effective size is not smaller than $\mathcal{L'}$.

Both systems can be described in terms of $2\times 2$ scattering matrix
$\hat S(\nu)$:
\begin{equation}
\label{eq:scatt_matrix}
\hat S(\nu)=\left( \begin{array}{cc} S_{11}(\nu)&S_{12}(\nu)\\
S_{21}(\nu)&S_{22}(\nu)\end{array} \right) \mbox{,}
\end{equation}
relating the amplitudes of the incoming and outgoing waves of
frequency $\nu$ in both leads. One should point out that for microwave systems it is customary to make measurements of the scattering matrices in a function of microwave frequency $\nu$ which is directly related to the real part of the wave number  $\mathrm{Re\,}k=\frac{2\pi }{c}\nu$.

In order to measure the two-port scattering matrix $\hat S(\nu)$ we connected the
vector network analyzer (VNA) Agilent E8364B to
the microwave networks shown in Fig.~1(c) and Fig.~1(d) and performed
measurements in the frequency range $\nu = 0.3-2.2$ GHz (see Fig.~2). The
connection of the VNA to a microwave network is equivalent to attaching of
two infinite leads $L^{\infty}_1$ and $L^{\infty}_2$ to quantum graphs which means that Fig.~1(a) and Fig.~1(b) correctly
describe the actual experimental arrangement.

 We considered two realizations of the Weyl and non-Weyl networks, W$_1$ and nW$_1$, and W$_2$ and nW$_2$, with the  lengths $ \mathcal{L}_1= 0.999$  m and $\mathcal{L}_2= 1.151$ m, respectively (see Table \ref{Lengths}).

\begin{table}
\begin{tabular}{ c c }
 \hline
 $ \mbox{Networks W$_1$ \& nW$_1$} $ & $\mbox{Networks W$_2$ \& nW$_2$} $\\
 \hline
 $\ell_1=0.127\pm 0.001\mbox{ m, }$ & $ \ell_1=0.203\pm 0.001\mbox{ m, }$\\
 \hline
 $\ell_2=0.103 \pm 0.001\mbox{ m, }$ & $ \ell_2=0.179 \pm 0.001\mbox{ m, }$\\
 \hline
 $\ell_3= 0.130 \pm 0.001 \mbox{ m, }$ & $ \ell_3=0.130 \pm 0.001 \mbox{ m, }$\\
 \hline
 $\ell_4=0.225 \pm 0.001 \mbox{ m, }$ & $ \ell_4=0.225 \pm 0.001 \mbox{ m, }$\\
 \hline
 $\ell_5=0.116\pm 0.001\mbox{ m, }$ & $ \ell_5=0.116\pm 0.001\mbox{ m, }$\\
 \hline
 $\ell_6=0.171\pm 0.001\mbox{ m, }$ & $ \ell_6=0.171\pm 0.001\mbox{ m, }$\\
 \hline
 $\ell_7=0.127\pm 0.001\mbox{ m, }$ & $ \ell_7=0.127\pm 0.001\mbox{ m. }$\\
 \hline
\end{tabular}
\caption{The optical lengths of the edges of the microwave  Weyl and non-Weyl networks, W$_1$ and nW$_1$, and W$_2$ and nW$_2$, with the lengths $ \mathcal{L}_1= 0.999$ m and $ \mathcal{L}_2= 1.151$ m, respectively. }
\label{Lengths}
\end{table}

   In the case of the non-Weyl networks nW$_1$ and nW$_2$, $\ell_2$ was the shortest edge  $\ell_s$ emanating from the balanced vertex of these networks, therefore, their effective sizes were $\mathcal{L'}_1= 0.896$ m and $\mathcal{L'}_2= 0.972$ m, respectively. The uncertainties in the edges' lengths of the networks are due to the preparation of coaxial microwave cables.

The moduli $|\det\bigr(\hat S(\nu)\bigl)|$ of the
determinants of the scattering matrices of the experimentally studied  Weyl and non-Weyl
networks in the frequency range 0.3 -- 2.2 GHz are shown in Fig.~2. Figs.~2(a) and 2(b) show the comparison of the experimental results obtained for the Weyl and non-Weyl  microwave networks W$_1$  and nW$_1$ containing $\ell_s=\ell_2=0.103$ m edge. Figs.~2(a) and 2(b) clearly show that the presence of the balanced vertex dramatically changes the spectrum of the non-Weyl network and lowers its number of resonances. For the Weyl network W$_1$ from the Weyl's formula ($\ref{eq:asym}$) in the frequency range 0.3 -- 2.2 GHz we should expect $N \simeq 13$ resonances while for the non-Weyl network nW$_1$ (formula ($\ref{eq:non_Weyl_asym}$)) $N \simeq 11$ resonances.  Indeed, in Fig.~2(a) we observe 13 resonances in agreement with the prediction of the Weyl's formula while in Fig.~2(b) we see only 11 resonances but in the agreement with the modified Weyl's formula ($\ref{eq:non_Weyl_asym}$).  In order to test more versatilely the applicability of the modified Weyl's formula ($\ref{eq:non_Weyl_asym}$) we compared our experimental results  with the theoretical ones (red arrows) obtained using the method of pseudo-orbits expansion.

\begin{figure}[tb]
\includegraphics[width=0.8\linewidth]{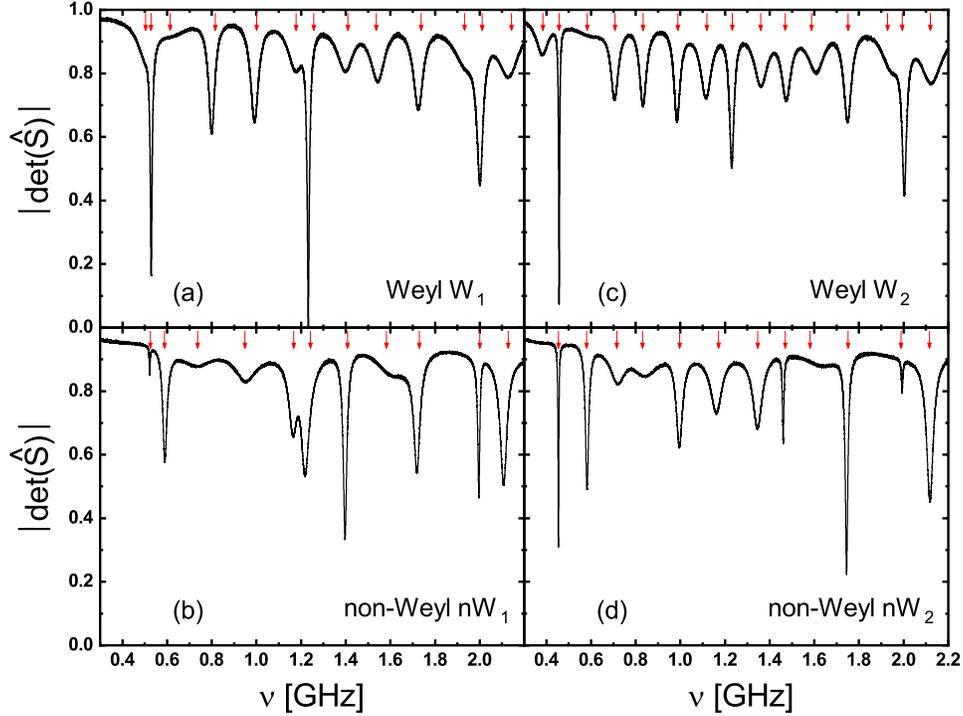}
\caption{
Panels (a) and (c) show the modulus of the
determinant of the scattering matrix  $|\det\bigr(\hat S(\nu)\bigl)|$  of the experimentally studied  Weyl
networks W$_1$ and W$_2$ (full lines) with the lengths $ \mathcal{L}_1= 0.999$ m and $ \mathcal{L}_2= 1.151$ m, respectively, containing two unbalanced vertices in the frequency range 0.3--2.2 GHz.
Panels (b) and (d) show the modulus of the
determinant of the scattering matrix  $|\det\bigr(\hat S(\nu)\bigl)|$  of the experimentally studied  non-Weyl
networks nW$_1$ and nW$_2$ with the same lengths as the networks W$_1$ and W$_2$ but possessing the effective sizes $\mathcal{L'}_1= 0.896$ m and $\mathcal{L'}_2= 0.972$ m, respectively,  in the same frequency range. The non-Weyl networks contained one balanced vertex.
The theoretical positions of the expected resonances are marked with red arrows.
}
\label{Fig2}
\end{figure}

\section{The resonance conditions}
The resonance conditions for the non-Weyl and Weyl graphs can be found using the method of pseudo-orbits \cite{Li6,Li7,BHJ}
\begin{equation}
\label{eq:res_cond}
  \mathrm{det\,}[\hat I_{2N} - \mathrm{exp\,(ik\hat L) \hat \Sigma}] = 0\,,
\end{equation}
where $\hat I_{2N}$ is $2N\times 2N$ identity matrix, $N$ is the number of the internal edges of the graph (in our case $N = 7$) and $\hat L = \mathrm{diag(\ell_1,\ell_2, \dots, \ell_7, \ell _1, \dots, \ell_7)}$, where $\ell_1 = \ell(1,4)$, $\ell_2 = \ell(1,2)$,  $\ell_3 = \ell(2,3)$, $\ell_4 = \ell(4,5)$, $\ell_5 = \ell(3,4)$, $\ell_6 = \ell(3,5)$, $\ell_7 = \ell(2,5)$.
   Here, it important to point out that for open systems resonances show up as poles \cite{Kottos2003} occurring at complex wave numbers $k_l=\frac{2\pi}{c}(\nu_l -i\Delta\nu_l)$, where $\nu_l$ and $2\Delta\nu_l$ are associated with the positions and the widths of resonances, respectively.
 The scattering matrix $\hat \Sigma$ for the non-Weyl and Weyl graphs  and the resonance conditions imposed by the Eq. (\ref{eq:res_cond})   is discussed in the details in the supplementary materials.

The calculations presented in Figs.~2(a) and ~2(b) (red arrows) show that in the frequency range 0.3 -- 2.2 GHz we should expect 13 and 11 resonances, respectively, which is in agreement with both the experiment and the number of resonances predicted by the formulas ($\ref{eq:asym}$) and ($\ref{eq:non_Weyl_asym}$). Therefore, we clearly see that the presence of the balanced vertex in  the non-Weyl  network nW$_1$ lowered  the number of resonances expected for the Weyl network W$_1$ by 2. It is worth pointing out that for both networks the theoretical positions of the resonances are in very good agreement with the experimental ones.

In Figs.~2(c) and 2(d) we show the comparison of the experimental results obtained for the Weyl and non-Weyl  microwave networks W$_2$ and nW$_2$ with the lengths $ \mathcal{L}_2= 1.151$ m containing $\ell_s=\ell_2=0.179$ m edge. They were constructed to be longer than the microwave networks W$_1$ and nW$_1$. In this case from the Weyl's formula ($\ref{eq:asym}$) we expect to have $N \simeq 15$ resonances while for the non-Weyl network nW$_2$ the number of the resonances should be $N \simeq 12$.  Indeed, in the experimental spectrum of the Weyl network W$_2$ 15 resonances are easily identified. In the case  of the non-Weyl network nW$_2$ the close inspection of the spectrum reveals 12 resonances in agreement with the formula ($\ref{eq:non_Weyl_asym}$). The positions of the theoretical resonances are marked with red arrows. Also in this case the theoretical calculations fully confirm the number of resonances found experimentally for both the Weyl and non-Weyl networks.  In the case of W$_2$ and nW$_2$ networks the presence of the balanced vertex in the non-Weyl network nW$_2$ lowered  the number of resonances expected for the Weyl network by 3.

In summary, we used microwave networks which simulate quantum graphs to show that there exist graphs which do not obey a standard Weyl's law. Such graphs are called \emph{non-Weyl} graphs. For standard coupling conditions we demonstrated that the transition from a \emph{Weyl} graph to a \emph{non-Weyl} graph occurs if  a balanced vertex is introduced.  Our experimental results are in excellent agreement with the numerical calculations yielding the resonances of the networks. They distinctly show  that the number of measured resonances may significantly depend on the way how the measured system is connected to the external world.

\section{Acknowledgements}
This work was supported in part by the National Science Centre Grant No. 2016/23/B/ST2/03979. J. L. was supported by the internal grant project ``Introduction to quantum mechanics on graphs'' of the University of Hradec Kr\'alov\'e.


\section*{References}

\begin{thebibliography}{DEL10}

\bibitem{Pauling36} L. Pauling, J. Chem. Phys. {\bf 4}, 673--677 (1936).

\bibitem{Ruedenberg53} K. Ruedenberg and C. Scherr, J. Chem. Phys. {\bf 21}, 1565--1581 (1953).

\bibitem{Exner88} P. Exner, P. \v{S}eba, and P. \v{S}\v{t}ov\'i\v{c}ek, J. Phys. A {\bf 21}, 4009--4019 (1988).

\bibitem{Kottos1997} T. Kottos and U. Smilansky, Phys. Rev. Lett. {\bf 79}, 4794 (1997).

\bibitem{Blumel2002} R. Bl\"umel, Yu Dabaghian, and R. V. Jensen, Phys. Rev. Lett. {\bf 88}, 044101 (2002).

\bibitem{BK} G. Berkolaiko and P. Kuchment, Introduction to Quantum Graphs (Mathematical Surveys and Monographs 186, 2013), p. 270.

\bibitem{Pluhar2014} Z. Pluha\v r and H. A. Weidenm\"uller, Phys. Rev. Lett. {\bf 112}, 144102 (2014).

\bibitem{Kowal1990} D. Kowal, U. Sivan, O. Entin-Wohlman, and Y. Imry, Phys. Rev. B {\bf 42}, 9009 (1990).

\bibitem{Imry1996} Y. Imry, {\it Introduction to Mesoscopic Systems} (Oxford, NY, 1996).

\bibitem{Sanchez1988} J. A. Sanchez-Gil, V. Freilikher, I. Yurkevich, and A.~A. Maradudin, Phys. Rev. Lett. {\bf 80}, 948 (1998).

\bibitem{Mittra1971} R. Mittra, S. W. Lee, {\it Analytical Techniques in the Theory of Guided Waves} (Macmillan, NY, 1971).

\bibitem{Hul2004} O.~Hul, S.~Bauch, P.~Pako\'nski, N.~Savytskyy, K.~{\.Z}yczkowski, and L.~Sirko, Phys. Rev. E {\bf 69}, 056205 (2004).

\bibitem{Sirko2016} M.  {\L}awniczak, S. Bauch, and L. Sirko, in Handbook of Applications of Chaos Theory,
 eds. Christos Skiadas and Charilaos Skiadas (CRC Press, Boca Raton, USA, 2016), p. 559.

\bibitem{Hul2012} O. Hul, M.~{\L}awniczak, S. Bauch, A. Sawicki, M. Ku\'s, L. Sirko, Phys. Rev. Lett {\bf 109}, 040402 (2012).

\bibitem{Bialous2016} M. Bia{\l}ous, V. Yunko, S. Bauch, M. {\L}awniczak, B. Dietz, and L. Sirko, Phys. Rev. Lett. {\bf 117}, 144101 (2016).

\bibitem{EL1} P. Exner and J. Lipovsk\'{y}, Equivalence of resolvent and scattering resonances on quantum graphs, Adventures in Mathematical Physics (Proceedings, Cergy-Pontoise 2006, {\bf 447}, Providence, R.I., 2007), pp. 73.

\bibitem{Li5} J. Lipovsk\'{y}, Acta Physica Slovaca {\bf 66}, 265 (2016).

\bibitem{DP} E.~B. Davies and A. Pushnitski, Analysis and PDE {\bf 4}, 729 (2011).

\bibitem{Weyl11} H. Weyl, Nachrichten der K\"oniglichen Gesellschaft der Wissenschaften zu G\"ottingen, 110--117 (1911).

\bibitem{DEL} E.~B. Davies, P. Exner and J. Lipovsk\'{y}, J. Phys. A: Math. Theor. {\bf 43}, 474013 (2010).

\bibitem{Li6} J. Lipovsk\'{y}, J. Phys. A: Math. Theor. {\bf 49}, 375202 (2016).

\bibitem{EL3} P. Exner and J. Lipovsk\'{y}, Phys. Lett. A {\bf 375}, 805 (2011).

\bibitem{Gnutzmann2006} S.~Gnutzmann and U.~Smilansky, Adv. Phys. {\bf 55}, 527 (2006).

\bibitem{Samuelson2004} K.A.~Dick, K.~Deppert, M.W.~Larsson, T.~M\"artensson, W.~Seifert, L.R.~Wallenberg,
and L.~Samuelson, Nature Mater. {\bf 3}, 380 (2004).

\bibitem{Heo2008} K.~Heo, E.~Cho, J.E.~Yang, M.H~Kim, M. Lee, B.Y.~Lee, S.G.~Kwon, M.S.~Lee, M.H.~Jo, H.J.~Choi, T.~Hyeon, and S.~Hong, Nano Lett. {\bf 8}, 4523 (2008).

\bibitem{Savytskyy2001} N. Savytskyy, A. Kohler, S. Bauch, R. Bl\"umel, and L. Sirko, Phys. Rev. E {\bf 64}, 036211 (2001).

\bibitem{Hul2005} O.~Hul, O.~Tymoshchuk, S.~Bauch, P.M.~Koch, and L.~Sirko, J. Phys. A {\bf 38}, 10489 (2005).

\bibitem{Lawniczak2008} M.~{\L}awniczak, O.~Hul, S.~Bauch, P.~\v Seba, and L.~Sirko, Phys. Rev. E {\bf 77}, 056210 (2008).

\bibitem{Lawniczak2010} M.~{\L}awniczak, S.~Bauch, O.~Hul, and L.~Sirko, Phys. Rev. E {\bf 81}, 046204 (2010).

\bibitem{Stockmann2016} A. Rehemanjiang, M. Allgaier, C.H. Joyner, S. M\"uller, M. Sieber, U. Kuhl, and H.-J. St\"ockmann, Phys. Rev. Lett. {\bf 117}, 064101 (2016).

\bibitem{Kottos2003} T. Kottos and U. Smilansky, J. Phys. A: Math. Gen. {\bf 36}, 350-1 (2003).

\bibitem{Dietz2017} B. Dietz, V. Yunko, M. Bia{\l}ous, S. Bauch, M. {\L}awniczak, and L. Sirko,  Phys. Rev. E {\bf 95}, 052202 (2017).

\bibitem{BHJ} R. Band, J.M. Harrison, C.H. Joyner, J. Phys. A Math. Theor. {\bf 45}, 325204 (2012).

\bibitem{Li7} J. Lipovsk\'{y}, Acta Physica Polonica A {\bf 128}, 968 (2015).


\end{thebibliography}
\end{document}